\documentclass[twocolumn,showpacs,showkeys,preprintnumbers,amsmath,amssymb]{revtex4}
\usepackage{graphicx}% Include figure files
\usepackage{amssymb,amsmath}
\usepackage{dcolumn}% Align table columns on decimal point
\usepackage{bm}% bold math
\begin{document}

\title{The antiferromagnetic insulator $ {\bf Ca_3FeRhO_6} $: characterization
       and electronic structure calculations}
\author{Volker Eyert}
\affiliation{TP II, Institut f\"ur Physik, Universit\"at Augsburg, D-86135
  Augsburg, Germany} 
\author{Udo Schwingenschl\"ogl} 
\affiliation{TP II, Institut f\"ur Physik, Universit\"at Augsburg, D-86135
  Augsburg, Germany} 
\author{Raymond Fr\'esard}
\affiliation{Laboratoire CRISMAT UMR CNRS-ENSICAEN 6508, 6 Boulevard
  Mar\'echal Juin, 14050 Caen Cedex, France} 
\author{Antoine Maignan}
\affiliation{Laboratoire CRISMAT UMR CNRS-ENSICAEN 6508, 6 Boulevard
  Mar\'echal Juin, 14050 Caen Cedex, France} 
\author{Christine Martin}
\affiliation{Laboratoire CRISMAT UMR CNRS-ENSICAEN 6508, 6 Boulevard
  Mar\'echal Juin, 14050 Caen Cedex, France} 
\author{Ninh Nguyen}
\affiliation{Laboratoire CRISMAT UMR CNRS-ENSICAEN 6508, 6 Boulevard
  Mar\'echal Juin, 14050 Caen Cedex, France} 
\author{Christian Hackenberger} 
\affiliation{EP VI, Center for Electronic Correlations and
  Magnetism, Institut f\"ur Physik, Universit\"at Augsburg, D-86135 Augsburg,
  Germany} 
\author{Thilo Kopp}
\affiliation{EP VI, Center for Electronic Correlations and
  Magnetism, Institut f\"ur Physik, Universit\"at Augsburg, D-86135 Augsburg, Germany}
\date{\today}
\begin{abstract}
We investigate the antiferromagnetic insulating nature of Ca$_3$FeRhO$_6$ 
both experimentally and theoretically. Susceptibility measurements
reveal a N\'eel temperature $ T_N \simeq 20 $\,K, and a magnetic 
moment of $5.3 \mu_B$/f. u., while M\"ossbauer spectroscopy strongly
suggests that the Fe ions,  located in trigonal prismatic sites, are in
a $3+$ high spin state. Transport measurements display a simple Arrhenius 
law, with an activation energy of $ \sim 0.2$\,eV. The experimental results 
are interpreted with LSDA band structure calculations, which confirm the 
$ {\rm Fe^{3+}} $ state, the high-spin/low-spin scenario, the 
antiferromagnetic ordering, and the value for the activation energy. 
\end{abstract}

\maketitle

\section{Introduction}

Interest in transition metal oxides has never been restricted to the 
most spectacular phenomenon of the high-$ T_c $ superconductivity, but 
also concerns, \emph{inter alia}, metal-insulator transitions, colossal 
magnetoresistance, and magnetic and orbital ordering \cite{Mae04}. Among 
the numerous magnetic transitions that have been studied, most of them 
are associated with structural transitions \cite{Rav99}. In contrast, 
no anomalous temperature dependence of the structural parameters has 
been reported in the $m = 0$, $n = 1$ members of the oxides family
A$_{3n+3m}$A'B$_{3m+n}$O$_{9m+6n}$, which are currently attracting much
attention \cite{Sti01}. These A$_3$A'BO$_6$ compounds crystallize in the
K$_4$CdCl$_6$ structure which consists of infinite chains along the c-axis
made of a 1~:~1 alternation of face-shared trigonal prisms
(A'O$_6$)$_{\mathrm{TP}}$ and octahedra (BO$_6$)$_{\mathrm{oct}}$. According
to their rhombohedral symmetry, the A cations separate the chains, the latter
forming a hexagonal array. For such compounds, when the A' trigonal
prism (TP) site is occupied by a magnetic cation, this provides
interesting physical properties created by the coexistence of
one-dimensionality character and geometrical frustration. This is illustrated
by 
Ca$_3$Co$_2$O$_6$ for which the ordered antiferromagnetic state below
$ T_N \sim 26 $\,K [\onlinecite{Fje96,Aas97}] bears some similarity 
to the partially disordered antiferromagnetic (PDA) state as originally 
proposed for ABX$_3$ geometrically frustrated 1D compounds \cite{Mek78}. 
But in marked contrast with the ABX$_3$ members, the intrachain coupling 
in Ca$_3$Co$_2$O$_6$ is ferromagnetic \cite{Fje96} and the magnetic 
field induced magnetization is very spectacular
\cite{Kag97,Kag97b,Mai00,Mai04}. Indeed, as a function of the applied magnetic
field, several magnetization jumps with a constant field spacing are
observed. Besides, the saturation magnetization is larger than 
expected from the assumption of different spin states for Co$^{3+}$ high
spin (HS) and low  spin (LS) in the TP and oct.,
respectively. Nonetheless, the ferromagnetic coupling along the chains is
probably related to this ``spin state ordering'', the latter resulting from
the different crystalline electrical fields in each Co$^{3+}$ polyhedron. Such
a coupling is likely to involve both LS  Co$^{3+}$ and O ions
\cite{Fre04,Eye04,Wu05}. In that respect, the different magnetic 
behavior of the two isostructural compounds Ca$_3$FeRhO$_6$ and
Ca$_3$CoRhO$_6$ is worth mentioning
\cite{Nii99,Dav03,Nii03,Nii01,Nii02,Har03,Nii01b,Loe03}. 
In the latter, the ferromagnetic intrachain coupling is expected as Rh$^{3+}$
is isoelectronic to Co$^{3+}$ (3d6), whereas the global magnetic behavior
of Ca$_3$FeRhO$_6$ appears to be antiferromagnetic although Fe$^{3+}$ (d5) or
Fe$^{2+}$ (d6) are both HS cations with large S values (5/2 or 2). 
Such different background states for the Ca$_3$MRhO$_6$ 1D compounds
(with M~=~Fe and M~=~Co) suggest subtle changes of the electronic  structure. 

In Ca$_3$FeRhO$_6$, contradicting results have been
reported for the oxidation states of iron and rhodium cations which
add more complexity to the interpretation. In order to shed light
on the magnetic and electronic behavior of Ca$_3$FeRhO$_6$, we compare, in
continuation of previous work \cite{Eye07}, results of band structure
calculations to the electrical and magnetic properties, together with
M\"ossbauer spectroscopy measurements.  

\section{Experiments}

The polycrystalline sample of Ca$_3$FeRhO$_6$ was prepared by mixing
the precursors CaO, Fe$_2$O$_3$ and Rh$_2$O$_3$ in the molar ratios
3~:~0.5~:~0.5. The thoroughly mixed powder, pressed in bars
($ {\rm \sim 2 \times 2 \times 10 mm^3} $) was first heated at 
$ 900^{\circ} $\,C for 24\,h and then at $ 1250^{\circ} $\,C for a 
$ 3 \times 24 $\,h period with intermediate X-ray controls.
The crystallinity and purity of the obtained black product were
checked by X-ray powder diffraction. The diffraction peaks have been
indexed in the space group R-3c with a and c values very close to
those reported in Refs.~[\onlinecite{Nii99,Dav03}], allowing to refine the 
3~:~1~:~1 ratio for the cation Ca~:~Fe~:~Rh with an uncertainty of
approximately 3~\% which is acceptable in all respects. Besides, small
intensity peaks were also found, that could be attributed to Ca$_2$Fe$_2$O$_5$.
Magnetic measurements were performed with a SQUID
magnetometer. Electrical resistivity was measured by the four probe
technique. The four electrical contacts were ultrasonically deposited
on a bar. The measurements were made by using a physical properties
measurement system (PPMS). The $^{57}$Fe powder M\"ossbauer resonance
spectrum at room temperature was performed with a transmission
geometry by use of a constant acceleration spectrometer and a
$\gamma$-ray source from $^{57}$Co embedded in a rhodium matrix. The
velocity scale was calibrated with an $\alpha$-Fe foil at room
temperature. The spectra were fitted with Lorentzian lines by the
unpublished MOSFIT program. The isomer shift was referred to metallic
$\alpha$-Fe at 293\,K.

\section{Results}
\subsection{Magnetism}

\begin{figure}[t!]
\begin{center}
\includegraphics*[width=0.45\textwidth]{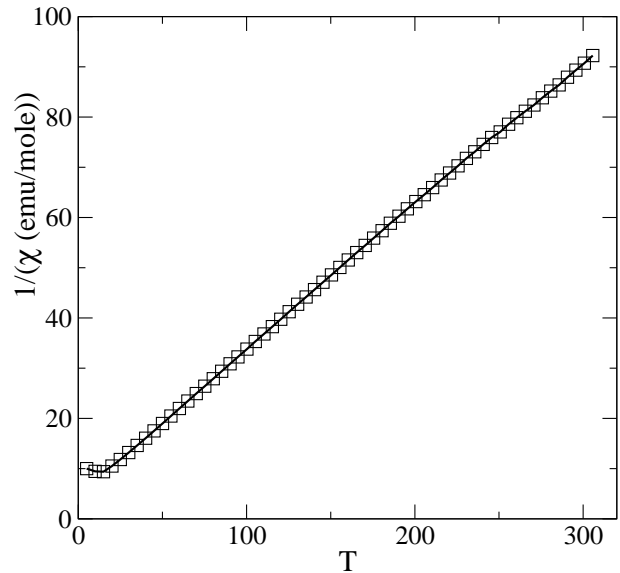}
\end{center}
\caption{Susceptibility of  Ca$_3$FeRhO$_6$. 
}
\label{fig:chi_FeRh}
\end{figure}
When compared to the $T$-dependent reciprocal magnetic susceptibility
curve [$\chi^{-1}(T)$] of  Ca$_3$Co$_2$O$_6$ as given in Ref.\ 
\cite{Mai00}, the $\chi^{-1}(T)$
curve of Ca$_3$FeRhO$_6$ (Fig.~\ref{fig:chi_FeRh}) exhibits a much
more linear behavior 
extending over a larger $T$ range. This result reflects a 
lack of ferromagnetic interactions for the latter as also attested by
the different extrapolated temperatures for
$\chi^{-1}(T=\theta_{\mathrm{CW}}) = 0$, the 
Curie-Weiss temperature $\theta_{\mathrm{CW}}$ values being $ -20 $\,K and
 $ +25 $\,K for 
Ca$_3$FeRhO$_6$ and Ca$_3$Co$_2$O$_6$, respectively. Furthermore in
Ca$_3$Co$_2$O$_6$, a $\chi^{-1}$
drop below $ T_N \sim 26 $\,K is observed, which indicates that a net
ferrimagnetic state is reached: on each triangle made by three
neighboring CoO$_6$ chains, two chains are antiferromagnetically coupled
(zero net magnetic moment) whereas the third one exhibits a net
ferromagnetic magnetization along the direction of the external applied
magnetic field. In contrast, the $\chi^{-1}(T)$ curve of
Ca$_3$FeRhO$_6$ exhibits a $\chi^{-1}$ increase below $ \sim 20 $\,K 
indicative of a 3D antiferromagnetic phase. Analysis of the slope along 
the linear region yields an effective paramagnetic moment 
$ \mu_{\mathrm{eff}}(\mathrm{exp.}) = 5.3 \mu_B/(Fe+Rh) $. 
We will refer to this experimental value below in subsection~C.  

The RT M\"ossbauer spectrum of this compound consists of a paramagnetic
doublet and is consistent with the measurements reported in Ref.~\cite{Nii03}. 
However, the best fit was obtained with two M\"ossbauer components, the 
hyperfine parameters of which are given in Table~\ref{table:Moss}. The 
observed isomer shift value ($ IS = 0.45 \pm 0.1 $\,mm/s) of the main 
component ($\% = 96 \pm 2$) is typical of Fe$^{3+}$ ions. Its high 
absolute quadrupole splitting value ($QS = 1.20 $\,mm/s) shows that this 
site is not in the octahedral symmetry, and therefore the Fe ions are 
located in trigonal sites. For the minor M\"ossbauer component 
($\% = 4 \pm 2$), the IS value of $ 0.73 \pm 0.1 $\,mm/s is rather 
corresponding to Fe$^{2+}$ ions. Therefore the vast majority of iron 
ions in Ca$_3$FeRhO$_6$ are in the trivalent state.  
\begin{table}[b!]
\caption 
{Refined $^{57}$Fe M\"ossbauer hyperfine parameters of Ca$_3$FeRhO$_6$
at room temperature including the linewidth $\Sigma$, and \% :
relative intensity of the M\"ossbauer site.} 
\label{table:Moss}
\begin{ruledtabular}
\begin{tabular}{cccc}
\multicolumn{1}{c}{IS (mm/s)} & 
\multicolumn{1}{c}{$\Sigma$ (mm/s)} & 
\multicolumn{1}{c}{QS (mm/s)} &
\multicolumn{1}{c}{\%} \\
\colrule
0.45 (1) & 0.28 (1) & 1.20 (1) & 96 (2)\\
0.73 (1) & 0.36 (1) & 1.49 (1) & 4 (2)\\
\end{tabular}
\end{ruledtabular}
\end{table}

\subsection{Transport} 

\begin{figure}[t!]
\begin{center}
\includegraphics*[width=0.45\textwidth]{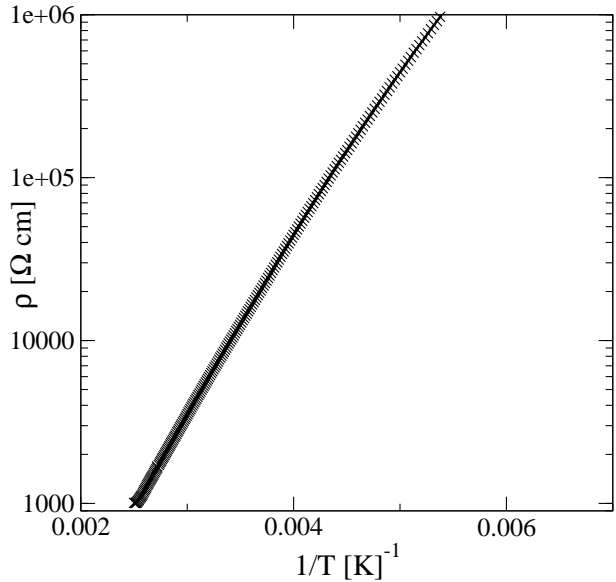}
\end{center}
\caption{Temperature dependence of the resistivity of
  Ca$_3$FeRhO$_6$, yielding an activation temperature $T_0 = 2200K$.}
\label{fig:rho_FeRh}
\end{figure}

The second set of measurements concerns the expected
localized nature of the electrical transport. Indeed, as shown in
Table~\ref{table1}, Ca$_3$FeRhO$_6$ appears far more insulating than
the related Co compounds since, for instance at 300\,K, the resistivity
$\rho$ for Ca$_3$FeRhO$_6$ is 160 times larger than that 
of Ca$_3$Co$_2$O$_6$ \cite{Mai03}. 
The $T$-dependence of the
resistivity  confirms that Ca$_3$FeRhO$_6$ is insulating, as shown in
Fig.~\ref{fig:rho_FeRh}. As $T$ decreases, $\rho$ increases
very rapidly in 
Ca$_3$FeRhO$_6$ reaching the set-up limit (corresponding to $\sim 10^6
\Omega$) at $\sim 230 $\,K. For the available $T$-range, the linear
$\ln{\rho} ~(T^{-1})$ curve shows that a  simple Arrhenius law is followed
from which an activation energy of 
0.2~eV can be extracted. According to both, high $\rho$ value and thermally
activated behavior, it turns out that in
Ca$_3$FeRhO$_6$, the charge carriers are
localized. Such a result 
is consistent with the antiferromagnetic intrachain coupling.  
These data for Ca$_3$FeRhO$_6$ confirm that despite the existing
similarities to isostructural Ca$_3$CoRhO$_6$, i.e., the A'$_{\mathrm{TP}}$ and
B$_{\mathrm{oct}}$ crystallographic sites are also occupied by trivalent
cations with high spin (S=5/2 for Fe$^{3+}$) and low spin (S=0 for Rh$^{3+}$),
respectively, the nature of the magnetic interactions differs
strongly.
\begin{table}[b!]
\caption 
{
Comparison of the resistivities at room temperature.
}
\label{table1}
\begin{ruledtabular}
\begin{tabular}{cccc}
\multicolumn{1}{c}{compound} & 
\multicolumn{1}{c}{Ca$_3$Co$_2$O$_6$} & 
\multicolumn{1}{c}{Ca$_3$CoRhO$_6$} &
\multicolumn{1}{c}{Ca$_3$FeRhO$_6$} \\
\colrule
{$\rho$(300\,K)  [$\Omega \cdot $ cm]}  & 50 [\onlinecite{Mai03}]& 39
[\onlinecite{Mai03}] &  8 300  \\
\end{tabular}
\end{ruledtabular}
\end{table}

\subsection{Band structure calculations} 

For the LSDA band structure calculations 
we used the augmented spherical wave (ASW) method in
its scalar-relativistic implementation \cite{wkg,aswrev}. In
the ASW method, the wave function is expanded in atom-centered
augmented spherical waves, which are Hankel functions and numerical
solutions of Schr\"odinger's equation, respectively, outside and inside
the so-called augmentation spheres. In order to optimize the basis set, 
additional augmented spherical waves were placed at carefully selected 
interstitial sites. The choice of these sites as well as the augmentation 
radii were automatically determined using the 
sphere-geometry optimization algorithm \cite{sgo}. The 
Brillouin zone integrations were performed using the linear tetrahedron 
method with up to 85 {\bf k}-points within the irreducible wedge.
In contrast to our previous work \cite{Eye07} we here use a new 
version of the ASW code, which takes the non-spherical contributions
to the charge density inside the atomic spheres into account.

\begin{figure}[htb]
\centering
\includegraphics[width=0.48\textwidth,clip]{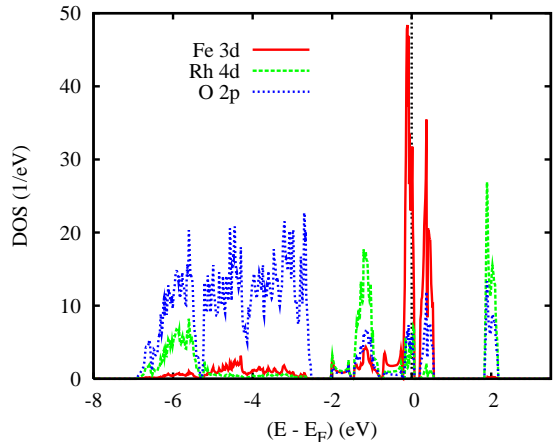}
\caption{(Color online) Partial densities of states (DOS) of spin degenerate 
         $ {\rm Ca_3FeRhO_6} $.}
\label{fig3}
\end{figure}

All calculations are based on the powder data of Niitaka {\em et al.} 
\cite{Nii99}. In a first step, we performed a set of calculations, 
where spin-degeneracy was enforced. The resulting partial densities 
of states (DOS) are shown in Fig.\ \ref{fig3}.
While O $ 2p $ dominated bands are located in the interval from 
$ -6.8 $ to $ -2.4 $\,eV, three groups of bands of mainly metal
$ d $-character are found at higher energies. However, the strong  
$ d $--$ p $ hybridization causes large $ p $/$ d $ contributions 
above/below $ -2 $\,eV, reaching up to 50\% especially for the Rh 
$ 4d $ states. 
 
\begin{figure}[htb]
\centering
\includegraphics[width=0.48\textwidth,clip]{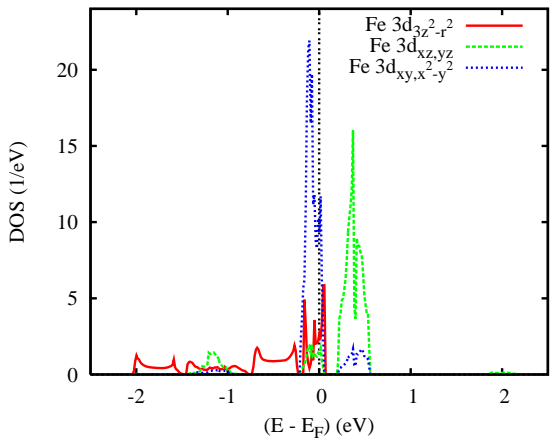}
\caption{(Color online) Partial Fe $ d $ DOS of spin degenerate $ {\rm Ca_3FeRhO_6} $.}
\label{fig4}
\end{figure}

\begin{figure}[htb]
\centering
\includegraphics[width=0.48\textwidth,clip]{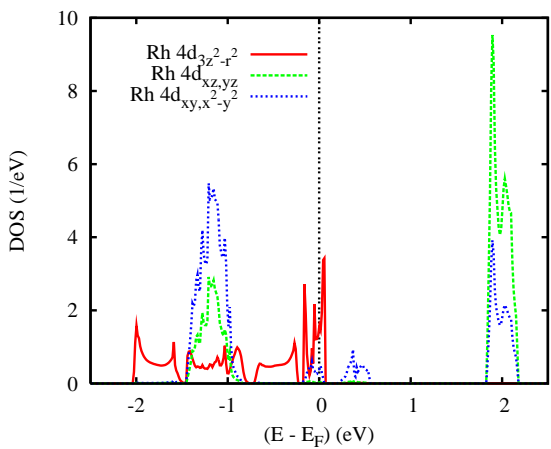}
\caption{(Color online) Partial Rh $ d $ DOS of spin degenerate $ {\rm Ca_3FeRhO_6} $.}
\label{fig5}
\end{figure}

According to the partial Fe $ 3d $ densities of states shown in Fig.\ 
\ref{fig4} 
the trigonal crystal field at the iron sites results in a splitting 
into non-degenerate $ d_{3z^2-r^2} $ as well as doubly degenerate 
$ d_{xy,x^2-y^2} $ and $ d_{xz,yz} $ states. The Rh $ 4d $ states 
as given in Fig.~\ref{fig5} 
experience a nearly perfect separation of the $ 4d $ states into 
occupied $ t_{2g} $ and empty $ e_g $ states due to the octahedral 
crystal field at these sites. Whereas strong $ \sigma $-type $ d $--$ p $
bonding places the Rh $ 4d $ $ e_g $ states at $ 2.0 $\,eV, the peak 
at about $ 0.4 $\,eV traces back to Fe $ d_{xz,yz} $ states. For this 
reason, spin-polarization of the latter bands is highly favourable,
with the observed high-spin/low-spin scenario.

In a second step, spin-polarized calculations were performed leading 
to the observed antiferromagnetic ordering, which is by 1\,mRyd per 
Fe atom more stable than the ferromagnetic configuration. Well localized 
magnetic moments of $ 0.00\,\mu_B $ (Rh), $ 3.77\,\mu_B $ (Fe), 
$ 0.13\,\mu_B $ (O), and $ 0.01\,\mu_B $ (Ca) are obtained in close 
agreement with those of previous calculations \cite{ville05}. These 
values reflect the experimental result of low- and high-spin states 
at the octahedral and trigonal prismatic sites, respectively. The 
total moment per sublattice amounts to $ \pm 4.58\,\mu_B $, which 
still might be slightly altered by the inclusion of spin-orbit coupling,  
which is beyond the present work. In particular, the obtained total 
magnetic moment per sublattice is smaller than the experimental value 
deduced from Fig.~\ref{fig:chi_FeRh} as was also observed for 
$ {\rm Ca_3Co_2O_6} $ \cite{Eye04,Wu05}. 

\begin{figure}[htb]
\centering
\includegraphics[width=0.48\textwidth,clip]{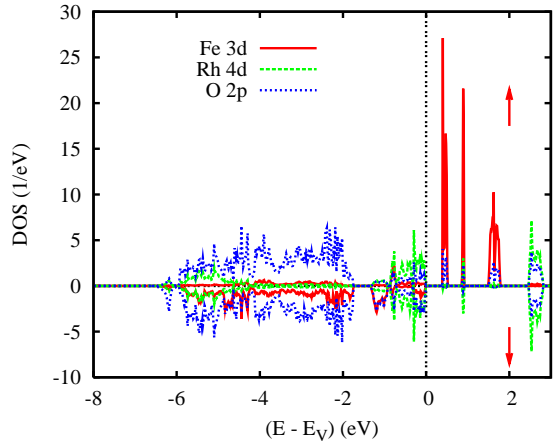}
\caption{(Color online) Partial DOS of antiferromagnetic $ {\rm Ca_3FeRhO_6} $.}
\label{fig6}
\end{figure}

Worth mentioning are the rather high magnetic moments at the oxygen sites 
arising from the strong $ d $--$ p $ hybridization, which sum up to 
about $ 0.8\,\mu_B $ per trigonal prism. Adding to the $ 3d $ moment 
they lead to the formation of extended localized moments already 
observed in $ {\rm Ca_3Co_2O_6} $ \cite{Eye04} and confirm the formal 
Fe $ S=5/2 $ configuration, hence, the formal $ {\rm Fe^{3+}} $ state. 
The high-spin behavior at the iron sites is clearly observed in the 
partial DOS shown in Fig.\ \ref{fig6},  
where the Fe $ 3d $ minority states display sharp peaks above $ E_F $ 
and the spin majority states are spread over a large energy interval 
as a result of the strong $ p $-$ d $ hybridization. 

The antiferromagnetic order growing out of the spin-polarized calculations 
goes along with the opening of an insulating gap of about 0.4\,eV as 
revealed by Fig.\ \ref{fig6}. This value corresponds to an activation 
energy of 0.2\,eV, which is in remarkably good agreement with the 
experimental value deduced from Fig.\ \ref{fig:rho_FeRh}.

\section{Summary}

In summary, we have performed susceptibility, M\"ossbauer spectroscopy,
and transport measurements on the antiferromagnetic insulating
compound $ {\rm Ca_3FeRhO_6} $. The experimental data have been 
compared with LSDA band structure calculations, and the agreement is 
found to be very good. In particular, the calculations confirm several 
experimental key results as, e.g., the charge and spin states at the 
Fe and Rh sites including the characteristic high-spin/low-spin scenario, 
the antiferromagnetic ordering, and the activation energy.

\section{Acknowledgements} 
This work is supported by the Deutsche Forschungsgemeinschaft through 
SFB 484.


\begin{thebibliography}{}

\bibitem{Mae04} S. Maekawa, T.~Tohyama, S.~E.~Barnes, S.~Ishihara,
  W.~Koshibae, and G. Khaliullin, {\it Physics of Transition Metal Oxides,} 
(Springer Verlag, Berlin (2004)).

\bibitem{Rav99} For a review, see {\it Colossal Magnetoresistance, Charge
Ordering and Related Properties of Manganese Oxides, }edited by C.\ N.\ R.\
Rao and B.\ Raveau (World Scientific, Singapore, 1998) and {\it Colossal
Magnetoresistance Oxides}, edited by Y.\ Tokura (Gordon \& Breach, London,
1999).
\bibitem{Sti01} K. E. Stitzer, J. Darriet, and H.-C. zur Loye,
  Curr. Opin. Solid State Mater. Sci. {\bf 5}, 535, (2001). 
\bibitem{Fje96} H. Fjellv\aa g, E. Gulbrandsen, S. Aasland, A. Olsen,
  and B. Hauback, J. Solid State Chem. {\bf 124}, 190 (1996). 
\bibitem{Aas97} S. Aasland, H. Fjellv\aa g, and B. Hauback, Solid
  State Comm. {\bf 101}, 187 (1997). 
\bibitem{Mek78} M. Mekata, J. Phys. Soc. Jpn. {\bf 42}, 76 (1977);
  M. Mekata and K. Adachi, J. Phys. Soc. Jpn, {\bf 44}, 806 (1978). 
\bibitem{Kag97} H. Kageyama, K. Yoshimura, K. Kosuge, H. Mitamura, and
  T. Goto, J. Phys. Soc. Jpn.  {\bf 66}, 1607 (1997). 
\bibitem{Kag97b} H. Kageyama, K. Yoshimura, K. Kosuge, M. Azuma,
  M. Takano, H. Mitamura, and T. Goto, J. Phys. Soc. Jpn.  {\bf 66}, 3996
  (1997). 
\bibitem{Mai00} A. Maignan, C. Michel, A.C. Masset, C. Martin, and
  B. Raveau, Eur. Phys. J. B {\bf 15}, 657 (2000). 
\bibitem{Mai04} A. Maignan, V. Hardy, S. H\'ebert, M. Drillon,
  M.R. Lees, O. Petrenko, D. McK. Paul, and D. Khomskii,
  J. Mater. Chem.  {\bf 14}, 1231 (2004). 
\bibitem{Fre04} R. Fr\'esard, C. Laschinger, T. Kopp, and V. Eyert,
  Phys. Rev B.  {\bf 69}, 140405(R) (2004). 
\bibitem{Eye04} V. Eyert, C. Laschinger, T. Kopp, and R. Fr\'esard,
  Chem. Phys. Lett.  {\bf 385}, 249 (2004). 
\bibitem{Wu05} H. Wu, M.W. Haverkort, Z. Hu, D. Khomskii, and
  L. H. Tjeng, Phys. Rev. Lett.  {\bf 95}, 186401 (2005). 
\bibitem{Nii99} S. Niitaka, H. Kageyama, M. Kato, K. Yoshimura and
  K. Kosuge, J. of Solid State  Chem.  {\bf 146}, 137, (1999).
\bibitem{Dav03} M.J. Davis, M.D. Smith and H.C. zur Loye, J. of Solid
  State Chem.  {\bf 173}, 122, (2003). 
\bibitem{Nii03} S. Niitaka, K. Yoshimura, K. Kosuge, K. Mibu,
  H. Mitamura, T. Goto, J. of Magn. and Magn. Mater.  {\bf 260}, 48, (2003). 
\bibitem{Nii01} S. Niitaka, K. Yoshimura, K. Kosuge, M. Nishi, and
  K. Kakurai, Phys. Rev. Lett.  {\bf 87}, 177202 (2001). 
\bibitem{Nii02} S. Niitaka, K. Yoshimura, K. Kosuge, A. Mitsuda,
  H. Mitamura, and T. Goto, J. Phys. Chem. Solids  {\bf 63}, 999 (2002). 
\bibitem{Har03} V. Hardy, M.R. Lees, A. Maignan, S. H\'ebert,
  D. Flahaut, and D. McK. Paul, J. Phys. Condens. Matt.  {\bf 15}, 5737
  (2003). 
\bibitem{Nii01b} S. Niitaka, H. Kageyama, K. Yoshimura, K. Kosuge,
  S. Kawano, N. Aso, A. Mitsuda, H. Mitamura, and T. Goto,
  J. Phys. Soc. Jpn.  {\bf 70}, 1222 (2001). 
\bibitem{Loe03} M. Loewenhaupt, W. Sch\"afer, A. Niazi, and
  E.V. Sampathkumaran, Europhys. Lett.  {\bf 63}, 374 (2003). 
\bibitem{Eye07}  V.~Eyert,
  U.~Schwingenschl\"ogl,  C.~Hackenberger, T.~Kopp, R.~Fr\'esard, and
  U.~Eckern,  Prog. Solid State
  Chem., at press (arXiv:cond-mat/0509374).

\bibitem{Mai03} A. Maignan, S. H\'ebert, C. Martin and D. Flahaut,
  Mater. Sc. and Eng. B  {\bf 104}, 121 (2003). 

\bibitem{wkg}
A.\ R.\ Williams, J.\ K\"ubler, and C.\ D.\ Gelatt, Jr.,
Phys.\ Rev.\ B {\bf 19}, 6094 (1979).

\bibitem{aswrev}
V.\ Eyert,
Int.\ J.\ Quantum Chem.\ {\bf 77}, 1007 (2000).

\bibitem{sgo}
V.\ Eyert and K.-H.\ H\"ock,
Phys.\ Rev.\ B {\bf 57}, 12727 (1998).

\bibitem{ville05}
A.\ Villesuzanne and M.-H.\ Whangbo, 
Inorg.\ Chem.\ {\bf 44}, 6339 (2005).




\end{thebibliography}
\end{document}